\documentclass[%
reprint,
amsmath,amssymb,
aps,
pra,
]{revtex4-2}

\usepackage{graphicx}
\usepackage{dcolumn}
\usepackage{bm}
\usepackage[utf8]{inputenc}
\usepackage{xcolor}
\usepackage{hyperref}

\begin{document}


\title{Static SERS with near-minus-one-epsilon substrate}
\author{Alexey P. Vinogradov}
\affiliation{Institute for Theoretical and Applied Electromagnetics, 125412, 13 Izhorskaya, Moscow, Russia}
 \affiliation{Dukhov Research Institute of Automatics (VNIIA), 22 Sushchevskaya, Moscow 127055, Russia;}
 \affiliation{Moscow Institute of Physics and Technology, 9 Institutskiy pereulok, Dolgoprudny 141700, Moscow region, Russia;}
\author{Evgeny S. Andrianov}
 \affiliation{Dukhov Research Institute of Automatics (VNIIA), 22 Sushchevskaya, Moscow 127055, Russia;}
 \affiliation{Moscow Institute of Physics and Technology, 9 Institutskiy pereulok, Dolgoprudny 141700, Moscow region, Russia;}
\affiliation{Institute for Theoretical and Applied Electromagnetics, 125412, 13 Izhorskaya, Moscow, Russia}

\date{\today}

\begin{abstract}
A mechanism for additional enhancement of SERS in the nanoparticle-on-mirror scheme is proposed.
This new mechanism is based on the use of a substrate made of material with a near-minus-one permittivity.
The setup involves a plasmonic nanoparticle in the form of an oblate ellipsoid positioned above the substrate and a Raman-active molecule located between them.
In the conventional  nanoparticle-on-mirror scheme, the plasmonic dipole resonance frequency coincides with the Stokes frequency of the Raman-active molecule.
Consequently, due to the Purcell effect, the molecule's near fields mostly excites a dipole mode in nanoparticle.
This dipole moment is many times greater than the dipole moment of the molecule by itself.
If the real part of substrate permittivity is near minus one, the image of the nanoparticle dipole moment in the mirror-substrate is a dipole moment pointed in the same direction but approximately $ 1/{\rm{Im}} \varepsilon_{_{\rm{ENZ}}}$  times larger in magnitude.
The simultaneous radiation of these two dipoles additionally increases the SERS intensity in $ 10^4$ times.
\end{abstract}

\maketitle

In 1928, C.V. Raman observed inelastic scattering of light by individual molecules~\cite{raman1928change}.
The observed inelastic scattering turned out to be much weaker than elastic, Rayleigh scattering.
To improve this, a rough substrate made of a noble metal is usually used~\cite{fleischmann1974raman,long2002raman}.
In the visible range from two to three eV, the real part of the permittivity of noble metals is negative.
For gold, it varies from $-10.0$ to $-1.7$, and for silver from $-17$ to $-3.8$~\cite{johnson1972optical} respectively.
On the rough substrate surface numerous nanoscale plasmonic structures with resonant properties in the visible wavelength range can be formed~\cite{le2024enhancement,lisyansky2024quantum}.
If the frequency of the incident field is close to the plasmon resonance frequency of the nanostructure, the field generated in this structure increases.
Raman scattering (RS) of a molecule, which is trapped in this structure, is enhanced~\cite{le2024enhancement}, and the effect of Surface Enhanced Raman Scattering (SERS) is observed.

Beyond rough noble metal substrates, artificially created structures have recently been used.
In such structures, thanks to their special geometry, there are regions where a strong enhancement of the electromagnetic field arises upon application of an external field.
These structures typically contain either metallic nanotips~\cite{zhang2013chemical} or narrow gaps between a plasmonic nanoparticle and a smooth substrate~\cite{lee2022nanoparticle}.
The latter scheme is referred to as nanoparticle-on-mirror scheme (NoMS).

\begin{figure}
\includegraphics[width=1\linewidth]{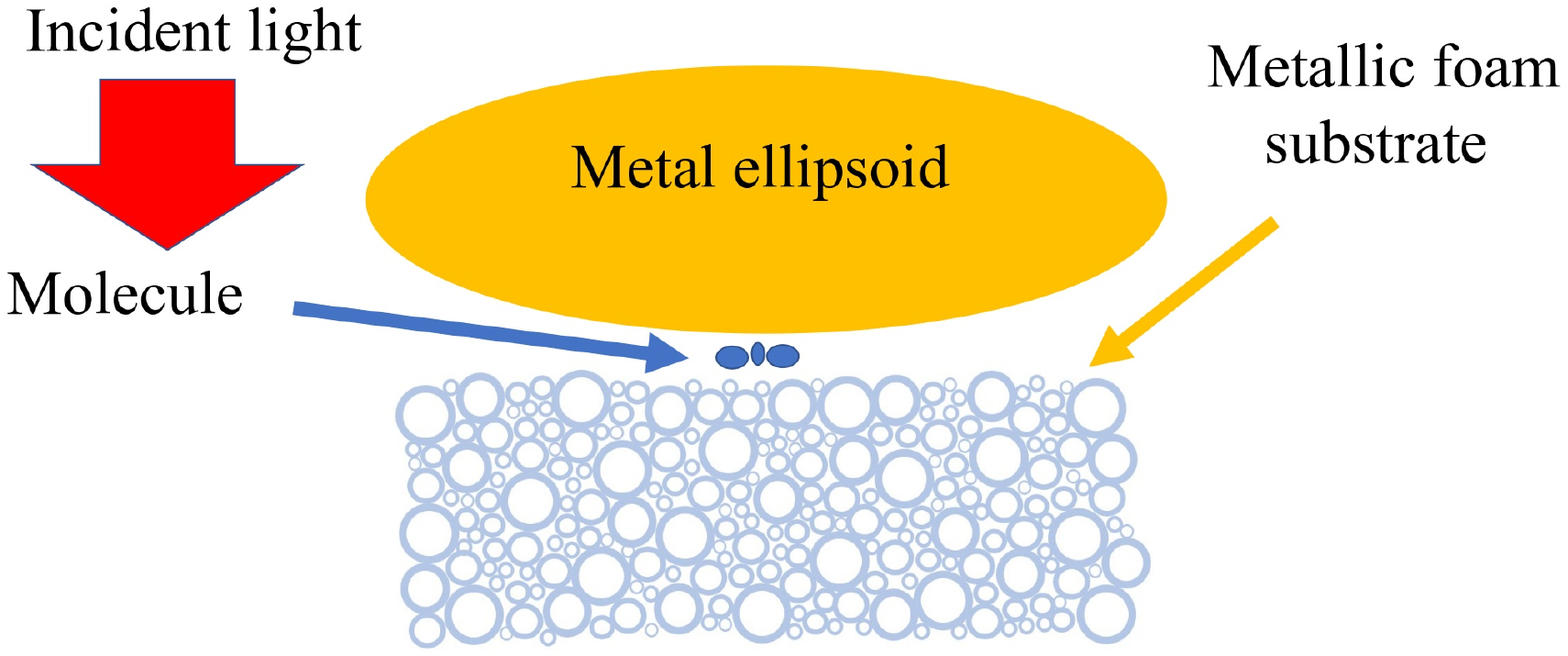}
\caption{ A scheme of an experiment with a substrate made of the HS medium, which consists of bubbles, all of which have the same ratio $r_{\rm{HS}}/R_{\rm{HS}} $ of the inner to the outer radius.}
\label{fig1}
\end{figure}

Below we propose an alternative, non-plasmonic mechanism for SERS.
To enhance Raman scattering, we suggest employing NoMS, in which a smooth substrate with a near-minus-one permittivity (NMOP) is used~(Fig.~\ref{fig1}). 

Note that under these conditions plasmon resonances are not excited~\cite{long2002raman,vinogradov2001compos}, and such substrates have not previously been used to observe Raman scattering, especially since for noble metals such values of the real part of the permittivity are observed only in the ultraviolet range of the spectrum, where observing Raman scattering is extremely difficult.
Currently, a search is underway for alternative plasmonic materials~\cite{beliaev2024alternative}.
The most promising are nitrides of transition metals, which have a permittivity near zero in the optical frequency range, but the imaginary part of their permittivity is of the order of several units, which hampers their applicability in the proposed scheme.

Nevertheless, in~\cite{liznev2010epsilon}, it was shown that a composite material with NMOP can be realized as noble metal foam for any visible light frequency.
The technology for obtaining such foam is proposed in~\cite{kuchmizhak2016fly}. 

From theoretical point of view such a composite material is the Hashin-Shtrikman (HS) medium, which entirely consists of metal bubbles (HS spheres) of various sizes much smaller than the wavelength, and completely filling the entire volume~\cite{milton2022theory,vinogradov2001compos}.
The latter means that all empty cavities among HS-spheres should filled with smaller HS-spheres~(Fig.~\ref{fig1}).
Below, we assume that the cavity of the HS sphere is filled with a vacuum. The ratio $x = r_{\rm{HS}}/R_{\rm{HS}}$  of the radius of the cavity of the HS sphere $r_{\rm{HS}} $  to its outer radius $R_{\rm{HS}} $ determines the metal concentration $p_{\rm{m}} $  in such a composite: $p_{\rm{m}} = 1 - x^3$.

In the quasi-static approximation, when the wavelength of incident wave is much greater than any characteristic dimensions of the problem, and the electric field is described by the Laplace equation, the effective permittivity of such a medium is determined from the condition that a HS sphere placed in a homogeneous medium with a permittivity equal to $\varepsilon_{\rm{HS}}$, does not disturb the homogeneous electric field applied on such a medium~\cite{milton2022theory,vinogradov2001compos}.
Since the Laplace equation does not have a characteristic length scale, this condition is simultaneously satisfied for all HS spheres smaller than the wavelength.
This self-consistency condition can be written as the HS equation~\cite{milton2022theory,vinogradov2001compos}
\begin{gather}
  \frac{1}{\varepsilon_{\rm{HS}} - 1} = \frac{1}{p_{\rm{m}}} \frac{1}{\varepsilon_{\rm{m}} - 1 } + \frac{1-p_{\rm{m}}}{3p_{\rm{m}}}
\end{gather}
Here $\varepsilon_{\rm{m}}$ is the permittivity value of the bulk metal, forming the shell of the HS-spheres.

Thus, at a demanded frequency the required permittivity of the composite ($-1 < {\rm{Re}} \varepsilon_{\rm{HS}} < 0 $) is achieved simply by selecting the concentration of the metal with a negative real part of the permittivity that forms the shell of HS-sphere, or more precisely by selecting the value of the parameter $x$. 

Note that the close packing of HS spheres of various sizes means that every point in this medium belongs to some HS sphere.
Recall that the polarization averaged over the volume of a HS sphere is independent of the sphere's size.
In other words, in this model, the local effective permittivity can be assumed to be free of fluctuations.
Theoretically, a smooth interface on which a molecule is located should consist of vanishingly small HS spheres, allowing the HS medium to be considered continuous.

An additional benefit of the HS medium is that at optical frequencies, losses in such a medium are much lower than those in solid metal. This is due to the fact that the dipole moment is determined by the size of the HS sphere, $R_{\rm{HS}}$, whereas the losses are determined by the volume of the shell.

The Raman-active molecule under study is positioned on a substrate beneath a nanoantenna, which is an oblate ellipsoid made of solid metal~(Fig.~\ref{fig1}).
This is the so-called nanoparticle-on-mirror arrangement.
The entire system is illuminated by an electromagnetic field of frequency $\omega_{\rm{ext}}$  incident normally to the substrate surface.
The wavelength of the incident field is significantly greater than all finite dimensions of the problem, including the nanoparticle size.

SERS is a two-step process.
Firstly, there should appear a place with enhanced local field.
Secondly, a Raman-emitting molecule must be a feeder for its immediate environment, which in turn must be an antenna.
Since Raman scattering is a linear in incident intensity processes both steps could be considered separately.
Below we focus our attention on the second step.

It is known that, in the case of the large Purcell effect~\cite{purcell1946spon}, a Raman-active molecule nonradiatively mostly transfers excitation to its plasmonic environment~\cite{bozhevolnyi2016fundamental}.
In our case, this is the metal ellipsoid and substrate, which then emit the received energy~\cite{bozhevolnyi2016fundamental}.
Clearly, the dipole moment of the ellipsoid (hereinafter referred to as a plasmonic nanoantenna (PNA)) is approximately $R_{\rm{PNA}}/R_{\rm{mol}}$  times greater than the dipole moment of the molecule~\cite{tamm1979fundamentals,smythe1987static}.
The ellipsoid should be chosen so that the frequency of its dipole mode frequency coincides with the Stokes frequency of the molecule.
Due to the Purcell effect, in the resonant case, the molecule transfers energy predominantly to the dipole mode of the PNA.

Since the size of the PNA is assumed to be much smaller than the wavelength, the field calculation can be performed in a quasi-static approximation.
To make some estimations we have to find, in addition to the dipole moment of PNA, its reflection in the mirror substrate.
According to the superposition principle, we can divide the charge distribution on a metallic nanoparticle into infinitesimally small charges.
We then have two options: find the potential of each charge and sum these potentials, or sum the charges of identical sign first and then determine the potential of resulting charges.
We follow the second way.
Employing the method of images, we calculate the potential of a charge $q$, which is located in upper half-space filled with vacuum, in the point with coordinates ${0,0,z_0}$.
The lower half-space is considered to be filled with a material with a permittivity $\varepsilon$. 

The potential in the upper half-space is searched as  $\varphi_1 = q/R + \varphi^\prime_1$ where $R = \sqrt{x^2 + y^2 + \left(z-z_0\right)^2 }$ is the distance from the observation point to the charge $q$.
The correction term  $\varphi^\prime_1$ satisfies the Laplace equation $\Delta \varphi^\prime_1 = 0$.
The potential $\varphi_2$ in the lower half space satisfies the Laplace equation $\Delta \varphi_2 = 0$.

On the surface $z = 0$ the potentials satisfy the boundary conditions 
\begin{gather}\label{cont}
  \left. \varphi_1 \right|_{z=0} = \left. \varphi_2 \right|_{z=0}
\end{gather}  
and 
\begin{gather}\label{divcont}
  \left. \frac{\partial \varphi_1}{ \partial z} \right|_{z=0} = \left. \frac{\partial \varphi_2}{\partial z} \right|_{z=0}
\end{gather}
The potentials depend on $R$  and $z$ or, for sake of convenience, on $R$  and $R^\prime = \sqrt{x^2 + y^2 + \left( z + z_0\right)^2 }$ that is equal to is the distance from the observation point to the image charge $q$.

Employing image method we search for correction term in the upper half-space as $\varphi_1 = \lambda q / R $  and the potential in the lower half-space as $\varphi_2 = \mu q / \varepsilon R $.
Eq.~(\ref{cont}) results in $1 + \lambda = \mu/\varepsilon$  whereas Eq.~(\ref{divcont}) yields $1 - \lambda = \mu$.
Finally, we arrive at the following expressions for the potentials~\cite{tamm1979fundamentals}:
\begin{gather}\label{potim}
  \varphi_1 \left({\bf{r}}\right) = \frac{q}{R} + \frac{1-\varepsilon}{1 + \varepsilon} \frac{q}{R^\prime}
\end{gather}

Thus, the image charge is located in the lower half-space, symmetrical to the real charge relative to the interface, at the distance $R^\prime$ from the interface.
Its charge is equal to $q\left(1 - \varepsilon \right)/\left( 1 + \varepsilon \right) $.
Due to the symmetry of the trial solution, the result does not depend on the choice of the merging point on the interface.

If $-1<\varepsilon<0$, the sign of this image charge coincides with the sign of the actual charge.
As $\varepsilon$  approaches minus one from above, the image charge can be $1/{\rm{Im}}\left(\varepsilon\left( \omega_0\right) \right)$ times greater than the actual charge, $q$.
In turn, the magnitude can reach $10^2$.

\begin{figure} 
\includegraphics[width=1\linewidth]{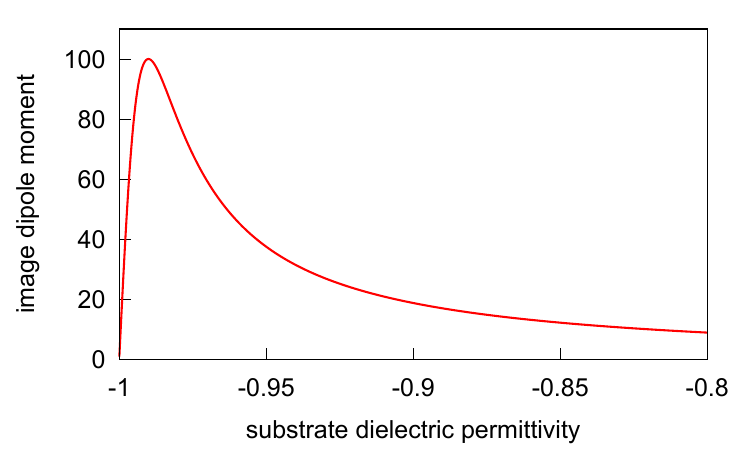}
\caption{
The real part of the complex dipole moment. The imaginary part is equal to $0.01i$.}
\label{fig2}
\end{figure}

The sign of this charge coincides with the sign of the real charge, and its magnitude, as $\varepsilon_{\rm{HS}}$ tends to minus one, can be $1/{\rm{Im}}\left(\varepsilon\left(\omega_0 \right) \right)$ times greater than the real charge (see~(Fig.~\ref{fig2})).
During the plasmonic oscillations of the PNA, its dipole moment determines the charges arising on its surface.
Similarly, the dipole moment of an image is created by the charges of these images.
The dipole moment of the image behaves similarly to the image of the charges that form it.
Its direction coincides with the direction of the dipole moment of the PNA, and the modulus increases in $\left(1 - \varepsilon_{\rm{HS}} \right)/\left( 1 + \varepsilon_{\rm{HS}} \right) $ times.
Consequently, according to the above estimate, the increase in intensity can reach $10^4$.
Note that this is an additional increase in the Raman signal observed in the NoMS~\cite{lee2022nanoparticle}.

In the presence of a substrate, to estimate the radiation intensity of whole system, it is necessary to consider not only the radiation of the PNA, but also the radiation from the substrate, which is equal to the radiation of the image of the dipole moment of the nanoantenna. 

It worth emphasizing the differences between the proposed scheme and the traditional NoMS~\cite{lee2022nanoparticle}.
The main difference is that the dipole moment should be directed not perpendicularly~\cite{zhang2013chemical,lee2022nanoparticle,krasnok2015antenna}, but parallel to the substrate surface.
Only in this case the fields from the dipole moment of the ellipsoid and from its image in the substrate add up, which increases the radiation intensity.
As a consequence, it is necessary to use an oblate ellipsoid, and the polarization of the incident field is chosen parallel to the substrate.
In such a scheme, normal field incidence must be used.
In the traditional scheme, $\varepsilon < -1 $, and the sign of the image is reversed.
In order for the dipole moments of the PNA and its image to be codirectional, the dipole moment of the molecule, which is parallel to the external field, must be directed perpendicular to the interface.
We also note that in this case the magnitude of the image charge does not increase as much, and the radiation intensity increases by no more than four times compared to the radiation from the molecule-on-nanoparticle system.

In conclusion, we note that in practice, an exact implementation of the Hashin-Shtrikman medium is impossible.
The Hashin-Shtrikman medium only allows for an analytical solution to the problem.
We believe that a metallic foam consisting of more or less identical bubbles may have the desired permittivity.
Of course, the question of the interaction of a nanoantenna with such a medium remains.
In general, introducing permittivity is always a phenomenological approximation.
One can suppose that if the nanoantenna size is an order of magnitude larger than the bubble size, then such an approach is justified.
It seems that all these problems are easy fix by experiment.
We consider our work as motivation for future experiments.

\bibliography{SERSMinOne}

\end{document}